\documentclass[a4paper,11pt]{amsart}
\usepackage{amscd,amsthm,amsfonts,latexsym,amssymb}
\usepackage[dvips]{graphicx}

\usepackage{bez123,calc,curves,ebezier,epic,eepic,graphicx,multiply,rotating}

\newtheorem{theorem}{Theorem} [section]

\newtheorem{proposition}[theorem]{Proposition}

\newtheorem{definition}[theorem]{Definition}

\newcommand{\norm}[1]{||} 

\newcommand{\dd}{{\mathrm d}}  

\newcommand{\RR}{{\bf R}}  
\newcommand{\CC}{{\bf C}}  

\newcommand{\ZZ}{{\bf Z}}  
\newcommand{\NN}{{\bf N}}  

\newcommand{\MM}{{\bf M}}  

\newcommand{\Ll}{{\mathcal L}}  
\newcommand{\Ss}{{\mathcal S}}  
\newcommand{\Nn}{{\mathcal N}}  

\newcommand{\CP}{{\bf C}P} 
\newcommand{\HP}{{\bf H}P} 

\newcommand{\ii}{{\rm i}}  
\newcommand{\jj}{{\rm j}}  

\newcommand{\ra}{\rightarrow}

\newcommand{\pa}{\partial}

\renewcommand{\phi}{\varphi}
\newcommand{\la}{\lambda}

\newcommand{\na}{\nabla}
\newcommand{\al}{\alpha}
\newcommand{\be}{\beta}

\newcommand{\Ga}{\Gamma}

\newcommand{\om}{\omega}
\newcommand{\Om}{\Omega}

\newcommand{\ov}{\overline} 

\newcommand{\SU}{\mbox{\rm SU}} 

\renewcommand{\Re}{{\rm Re}\,} 
\newcommand{\grad}{\mbox{\rm grad}\,}

\begin{document}

\title{Twistor theory on a finite graph}

\author{Paul Baird and Mohammad Wehbe}

\address{D\'epartement de Math\'ematiques,
Universit\'e de Bretagne Occidentale,
6 av.\ Victor Le Gorgeu -- CS 93837,
29238 Brest Cedex, France  }

\email{Paul.Baird@univ-brest.fr, Mohammad.Wehbe@univ-brest.fr}

\begin{abstract}
We show how the description of a shear-free ray congruence in Minkowski space as an evolving family of semi-conformal mappings can naturally be formulated on a finite graph.  For this, we introduce the notion of holomorphic function on a graph.  On a regular coloured graph of degree three, we recover the space-time picture.  In the spirit of twistor theory, where a light ray is the more fundamental object from which space-time points should be derived, the line graph, whose points are the edges of the original graph, should be considered as the basic object.  The Penrose twistor correspondence is discussed in this context.
\end{abstract}

\maketitle

\section{Introduction} 

Two appealing ideas, both due to R. Penrose, provide a different perspective to our understanding of physical fields.  The first of these is to try to build up space-time and quantum mechanics from combinatorial principles.  One way to attempt this is from so-called spin networks and Penrose argues how $3$-dimensional space arises from systems with large angular momentum.  A spin network is a graph whose vertices have degree $3$ (the number of edges incident with each vertex is $3$) and whose edges are labeled by an integer which represents twice the angular momentum \cite{Pe-4}.  The second idea is to consider twistor space, the space of null geodesics, as the more basic object from which space-time points should be derived \cite{Pe-3}.  Twistor diagrams can be considered as a natural adaptation of the combinatorial perspective to the twistor program \cite{Pe-3}. Our principal aim in this article is to give an alternative way in which combinatorial structures arise from the twistorial construction of fields. 

One of the basic objects of twistor theory, a shear-free ray congruence, can be viewed alternatively as a semi-conformal complex valued mapping which evolves in time \cite{Ba-We-1, Ba-Wo-1, Ba-Wo-2}.  This latter object, which we shall consider as a physical field, is perfectly suited to be defined on a finite graph (or network).  In this context, we shall refer to the function as \emph{holomorphic}, since in the plane, a semi-conformal mapping is either holomorphic or anti-holomorphic.  What is fascinating is that only certain graphs support a holomorphic function.  If the order of the graph is sufficiently small, computer programs can be used to generate such functions.

To a graph $\Ga$ endowed with a holomorphic function $\phi : V(\Ga ) \ra \CC$, where $V(\Ga )$ is the set of vertices of $\Ga$, we can associate its twistor dual $L_{\Ga}$, whose vertices are the edges of the original graph, sometimes called the \emph{line-graph}, as well as a function $\psi : V(L_{\Ga}) \ra \CC$.  In the spirit of twistor theory, where light rays are considered to be the fundamental objects, we consider the graph $L_{\Ga}$ as the basic object from which physical fields and space-time points should be deduced.  Indeed, a vertex of $\Ga$ arises as a complete subgraph in $L_{\Ga}$ upon which $\sum \psi^2$ vanishes.  An outline of the paper is as follows.

We first of all explain how a shear-free ray congruence on $4$-dimensional Minkowski space can be viewed as an evolving family of complex-valued semi-conformal mappings on $3$-dimensional space-like slices.  This is the basis of our generalization to graphs.  

In Section \ref{sec:graphs}, we discuss finite graphs.  In particular, we recall the notion of holomorphic mapping between graphs and introduce the concept of holomorphic function on a graph.  Holomorphic mappings are then characterized by the property that they preserve holomorphic functions (Proposition \ref{prop:holo}).  The properties of holomorphic functions are discussed in relation to quantum graphs, spin networks and orthographic projection.

A holomorphic function on a graph is equivalent to an isotropic $1$-form which vanishes around closed cycles.  On a regular graph of degree $3$ oriented by colour, we show how, from an isotropic $1$-form, we can recover a spinor field defined on the vertices, which corresponds to the spinor field defining a shear-free ray congruence on space-time.  Finally, in Section \ref{sec:line-graph}, we discuss the twistor correspondence between a graph and its line-graph.

\section{Shear-free ray congruences on Minkowski space} \label{sec:SFR}

The Penrose twistor correspondance associates to a light ray in Minkowsi space $\MM^4$,  a point in a $5$-dimensional CR-submanifold $\Nn^5$ of $\CP^3$ \cite{Pe-1}.  We can obtain $\Nn^5$ as follows.  

We first of all compactify $\MM^4$ by adding a light cone at infinity to obtain the manifold $\ov{\MM}^4$ diffeomorphic to $S^1 \times S^3$.  
The Hopf fibration $\pi : \CP^3 \ra S^4$ is the map given by
$$
\pi ([z_1, z_2, z_3, z_4]) = [z_1 + z_2 \jj , z_3 + z_4 \jj ] \in \HP^1
$$
where we use homogeneous coordinates $[z_1, z_2, z_3, z_4]$ for points of $\CP^3$ and where $\HP^1$ is the quaternionic projective space.  On identifying $\HP^1$ with $S^4$ and letting $S^3$ be the equatorial $3$-sphere given by $\Re [z_1 + z_2 \jj , z_3 + z_4 z\jj ] = 0$, we see that $\pi ([z_1, z_2, z_3, z_4]) \in S^3$ if and only if 
$$
z_1 \ov{z_3} + z_2 \ov{z_4} + \ov{z_1} z_3 + \ov{z_2}z_4 = 0\,.
$$
We then define $\Nn^5 = \pi^{-1}(S^3)$.  Note that $S^3$ is parallizable and so the bundle $\Nn^5$ is diffeomorphic with $S^3 \times S^2$.  There is now a natural identification between $\pi^{-1}(S^3)$ and the unit tangent bundle $T^1S^3$ to $S^3$.  If we consider $S^3$ as the compactified slice $t = 0$ in $\ov{\MM}^4$, then a point $(x,v)$ of $\Nn^5 \cong S^3 \times S^2$ gives the light ray passing through $x$ with direction $v$.      

The picture can be unified by introducing the flag manifold ${\bf F}_{12}$ of pairs $(\ell , \Pi )$ consisting of resp. $1$- and $2$-dimensional subspaces of $\CC^4$ with $\ell \subset \Pi$ and considering the double fibration:
$$
\begin{array}{ccccc}
 & & {\bf F}_{12} & & \\
 & \swarrow & & \searrow & \\
\CP^3 & & & & G_2(\CC^4) \\
\cup & & & & \cup \\
\Nn^5 & & & & \ov{\MM}^4
\end{array}
$$ 
where $G_2(\CC^4)$ is the Grassmannian of complex $2$-dimensional subspaces of $\CC^4$ and where the left projection is given by $(\ell , \Pi ) \mapsto \ell$ and the right by $(\ell , \Pi ) \mapsto \Pi$.  A point $\ell$ of $\CP^3$ determines a plane in $G_2(\CC^4)$, called an $\al$-plane, given by all the $\Pi$ containing $\ell$.  This plane may or may not intersect $\ov{\MM}^4$; if it does it does so in a null geodesic.  The points of $\CP^3$ which gives light rays are precisely the points of $\Nn^5$.  

In order to describe a shear-free ray congruence (SFR), it is useful to have the notion of conformal foliation.  We formulate this in terms of a semi-conformal mapping, which will be the fundamental object we discuss later in the context of graphs.  

A Lipschitz map $\phi : (M^m, g) \ra (N^n, h)$ between Riemannian manifolds is said to be \emph{semi-conformal} if, at each point $x\in M$ where $\phi$ is differentiable (dense by Radmacher's Theorem), the derivative $\dd \phi_x : T_xM\ra T_{\phi (x)}N$ is either the zero map or is conformal  and surjective on the complement of $\ker \dd\phi_x$ (called the \emph{horizontal distribution}).  Thus, there exists a number $\la (x)$ (defined almost everywhere), called the \emph{dilation}, such that $\la (x)^2 g(X,Y)$ $ = $ $\phi^*h(X,Y)$, for all $X,Y\in (\ker \dd\phi_x)^{\perp}$.  If $\phi$ is of class $C^1$, then we have a useful characterisation in local coordinates, given by
$$
g^{ij}\phi_i^{\al}\phi_j^{\be} = \la^2 h^{\al\be}\,,
$$ 
 where $(x^i), (y^{\al})$ are coordinates on $M, N$, respectively and $\phi^{\al}_i = \pa (y^{\al}\circ \phi ) / \pa x^i$.  The  fibres of a smooth submersive semi-conformal map determine a conformal foliation, see \cite{Va} and conversely, with respect to a local foliated chart, we may put a conformal structure on the leaf space with respect to which the projection is a semi-conformal map.  We then have the identity:
$$
\left(\Ll_Ug\right)(X,Y) = - 2 U(\ln \la )\,g(X,Y),
$$
for $U$ tangent and $X,Y$ orthogonal to the foliation.  This latter equation can be taken to be the characterisation of a conformal foliation.  Specifically, a foliation is called \emph{conformal} if there is a function $a = a(U)$ which depends only on $U$, such that 
$$
\left(\Ll_Ug\right)(X,Y) = a(U) \,g(X,Y),
$$
for $U$ tangent and $X,Y$ orthogonal to the foliation.  The relation between $a$ and the dilation $\la$ can now be deduced by calculating the mean curvature of the horizontal distribution, see, for example \cite{Ba}.  

A shear-free ray congruence on a region $A \subset \MM^4$ is a foliation by null-geodesics which is without shear.  That is, if $W$ represents the tangent vector field to the congruence, then at a point $x\in A$, the metric complement $W^{\bot}$ is $3$ dimensional and contains $W$ itself; if we take a $2$-dimensional spacelike complement $S$ in $W^{\bot}$, then for the congruence to be \emph{shear-free}, we require Lie transport of vectors in $S$ along $W$ to be conformal.  This property is independent of the choice of $S$.  By the Kerr Theorem, locally a shear-free ray congruence is defined by the intersection of $\Nn^5$ with a complex analytic surface $\Ss$ \cite{Pe-Ri-2}.  In general the congruence of light rays defined by $\Nn^5 \cap \Ss$ will be multivalued with singularities.  Solutions of the zero rest-mass field equations
$$
\na^{AA'}\phi_{AB\ldots L} = 0
$$
are then given by considering a function $f(z_1, z_2, z_3, z_4)$ homogeneous of degree $- n - 2$ and taking a contour integral in an appropriate way.  This is the basis of the Penrose transform, which is an integral transform from sheaf cohomology in the twistor space into the space of massless fields, see \cite{Ea, Pe-3, Wa-We} for details.

In \cite{Ba-We-1}, the equations for an SFR are reformulated in such a way that will enable us to adapt them to the context of graphs.  Specifically, if $W$ is tangent to a future pointing congruence of null curves on a region $A \subset \MM^4$, then at each point $(t,x) \in \MM^4$, we can decompose $W$ into its timelike and spacelike components: $W = \pa_t + U$, where $U$ is a unit tangent to the slice $\RR^3_t = \{ (t, x_1, x_2, x_3) \in \MM^4: t \ {\rm const}\ \}$.  Then $W$ is tangent to an SFR if and only if 
\begin{equation} \label{SFR}
\left\{ \begin{array}{lrcl}
{\rm (i)} &  \frac{\pa U}{\pa t} & = & - \na^{\RR^3_t}_UU \\
{\rm (ii)} & 0 & = & (\Ll_Ug) (X + \ii Y,X + \ii Y) \,,
\end{array} \right.
\end{equation}
where $\{ X, Y, U \}$ is an orthonormal basis tangent to $\RR^3_t$ at each point and $g$ is the standard Euclidean metric on $\RR^3_t$.   Indeed, the unit direction field $U$ can be represented by a spinor field $[\mu^A]\in \CP^1$ and then (\ref{SFR}) is equivalent to the usual spinor representation of an SFR:
$$
\mu^A\mu^B\na_{AA'}\mu_B = 0\,.
$$
Note that (\ref{SFR})(i) is equivalent to the geodesic condition $\na^{\MM^4}_WW = 0$, whereas (\ref{SFR})(ii) is equivalent to the property that $U$ be tangent to a conformal foliation on each slice $\RR^3_t$.  Furthermore, one can show that if (\ref{SFR})(i) is satisfied everywhere and (\ref{SFR})(ii) on an initial slice $\RR^3_0$, then (\ref{SFR})(ii) is satisfied for all $t$ \cite{Ba-We-1}. 

If we locally integrate the vector field $U$, so that for each $t$ it is tangent to the fibres of a semi-conformal mapping $\phi = \phi_t: B_t  \ra \CC$ ($B_t$ open in $\RR^3_t$),  the above equations are equivalent to the pair \cite{Ba-We-1}:
\begin{equation} \label{evol-scm}
\left\{ \begin{array}{lrcl}
{\rm (i)} & d\left( \frac{\pa \phi }{\pa t} \right) (U) & = & -\tau (\phi )  \\
{\rm (ii)} & 0 & = & g(\grad \phi , \grad \phi ) \,,
\end{array} \right.
\end{equation}
where $\grad \phi$ is the (complex) gradient with respect to the metric $g$ on $\RR^3_t$.
In fact one can easily check that \eqref{evol-scm} is invariant under the replacement of $\phi_t$ by $\psi_t = \zeta_t \circ \phi_t$, where $\zeta_t$ is an arbitrary conformal transformation of a domain of the complex plane; this is precisely the gauge freedom one requires in the choice of $\phi_t$.

\section{Holomorphic functions on a graph}  \label{sec:graphs}

A \emph{finite graph $\Ga$ of order $n$} is a set $V$ of cardinality $n$ endowed with a binary relation $\sim$.  For $x,y \in V$, if $x\sim y$ we will say that $x$ and $y$ are \emph{neighbours}, or are \emph{joined by an edge} and we will represent this diagrammatically by drawing a line segement between $x$ and $y$.  We suppose in what follows that the relation $\sim$ is symmetric, so that edges are not directed, although most of our discussion also applies to directed graphs.  We prefer to use the term \emph{directed}, rather than the more usual \emph{oriented}; the latter term being reserved for a notion of orientation of a (undirected) graph, rather akin to orientation of a manifold, which we will define later.  We do not allow the relation $\sim$ to be reflexive, so that the graph $\Ga$ does not contain loops, neither do we allow multiple edges, although once more, the discussion can be adapted to this more general situation.  We can represent the edges as a subset $E$ of the formal symmetric product $V\odot V$ and so express the graph $\Ga$ as the pair $\Ga = (V,E)$.   It will often be convenient to represent an edge $(x,y)\in E$ using the notation $\ov{xy}$, or, if we impose a direction on the edge, by $\vec{xy}$.  We say that the edge $\ov{xy}$ is \emph{incident with the vertex $x$} (and also with $y$).    

It is our aim to represent fields purely in terms of the combinatorial properies of graphs and as far as possible to dispense with notions of (semi-) Riemannian geometry.  However, a natural generalisation of our theory is to endow each edge with a real number, called its \emph{length} and to consider what are called \emph{metric graphs}.  One can even go further, and suppose that an angle is defined between edges incident with a given vertex, however, this now becomes an approximation of (semi-)Riemannian geometry and would defeat our purpose of developing a purely \emph{combinatorial} theory.

Many notions of Riemannian geometry translate into combinatorial properties of graphs.  A useful references is the book by Chung \cite{Ch}, which uses slightly different conventions.  We outline below, those notions which are essential to our development.

Given a graph $\Ga = (V, E)$, to each $x \in V$, we define its \emph{degree} $m(x)$ to be the number of edges incident with $x$.  A graph is called \emph{regular} if $m(x) = m$ is constant for each vertex.  We define the \emph{tangent space at $x\in V$}, to be the set $T_x\Ga := \{ \vec{xy}: \ov{xy}\in E\}$.  That is, each element of $T_x\Ga$ is a \emph{directed} edge, with base point $x$ and end point $y\sim x$.  Given a function $\phi : V \ra \RR^N$ with values in a Euclidean space and a vector $X = \vec{xy} \in T_x\Ga$, we define its \emph{directional derivative in the direction} $X$ to be the number
$$
d\phi_x(X) = \phi (y) - \phi (x)\,.
$$
Note that we could extend the notion of tangent space to include all linear combinations of edges $\vec{xy}$, $y\sim x$, to obtain a vector space, but we prefer to use a discrete concept for the tangent space.  If $\om : T_x\Ga \ra \RR^N$, then we define its \emph{co-derivative at $x$} to be the quantity
$$
\dd^* \om (x) = -\frac{1}{m(x)}\sum_{y\sim x} \om (\vec{xy})\,.
$$
If for each $x\in V$ we have given a map $\om = \om_x : T_x\Ga \ra \RR^N$, then provided $\om (\vec{xy}) = - \om (\vec{yx})$, we will refer to $\om$ as an $\RR^N$-valued $1$-form.  In particular, if $f : V \ra \RR$ is a function, then $\dd f$ is a $1$-form and we have
\begin{eqnarray*}
\dd^*\dd f (x) & = & - \frac{1}{m(x)}\sum_{y\sim x} (f (y) - f (x)) \\
  & = &  f (x) - \frac{1}{m(x)}\sum_{y \sim x} f (y) \\
  & = & \Delta f (x)\,,
\end{eqnarray*}
where we define the \emph{Laplacian of $f$} to be the quantity
$$
\Delta f (x) := f (x) - \frac{1}{m(x)} \sum_{y\sim x} f (y)\,.
$$
Note that our sign convention for the Laplacian is such that its eigenvalues are positive.  

The notion of semi-conformal mapping between graphs was introduced by H. Urakawa in 2000 \cite{Ur-1, Ur-2}.  More recently, these have been called \emph{holomorphic mappings} by M. Baker and S. Norine in their development of Riemann surface theory in the context of finite graphs \cite{Ba-No-1, Ba-No-2}.  Motivated by our Proposition \ref{prop:holo} below, we shall also refer to these as holomorphic mappings between graphs.  

Let $\Ga_1 = (V_1,E_1)$ and $\Ga_2 = (V_2, E_2)$ be two (not necessarily finite) graphs.  Then a mapping $\phi : V_1 \ra V_2$ between the vertices is defined to be a \emph{mapping of graphs}, if, whenever $x\sim y$ $(x,y\in V_1$) we have, \emph{either} $\phi (x) = \phi (y)$, \emph{or} $\phi (x) \sim \phi (y)$.  In this case we will write: $\phi : \Ga_1 \ra \Ga_2$.   

\begin{definition} \label{def:semi-conf}  Let $\phi : \Ga_1 = (V_1, E_1) \ra \Ga_2 = (V_2, E_2)$ be a mapping of graphs.  Then we say that $\phi$ is {\rm holomorphic at} $x\in V_1$ if, on setting $z = \phi (x)$, for all $z'\sim z$, the number 
$$
\la (x, z'):= \sharp \{ x'\sim x : \phi (x') = z'\}\,,
$$
is well-defined and depends only on $x$ (i.e. it is independent of the choice of $z'$) in which case we write $\la (x) = \la (x, z')$.  We say that $\phi$ is {\rm holomorphic} if it is holomorphic at every point.  In this case, if $x\in V_1$ is such that $\phi (y) = \phi (x)$ for all $y\sim x$, we set $\la (x) = 0$ and so obtain a well-defined function $\la : V_1 \ra \NN$, called the {\rm dilation} of $\phi$.
\end{definition}

The above definition can easily be extended to mappings of metric graphs, where now $\Ga_1$ and $\Ga_2$ are endowed length functions $\ell_1, \ell_2$ defined on the edges $E_1, E_2$, respectively \cite{An}.  The dilation is then replaced by the function
$$
\la (x) = \ell_2(\ov{\phi (x)z'})\sum_{\stackrel{x'\sim x}{\phi (x') = z'}}\frac{1}{\ell_1(\ov{xx'})}\,.
$$

An \emph{automorphism} of a graph $\Ga = (V,E)$ is a bijective mapping $\phi : V \ra V$ such that $x\sim y$ if and only if $\phi (x) \sim \phi (y)$.  It follows that an automorphism is holomorphic with dilation identically equal to $1$.  We interpret such a mapping as the analogue of an \emph{isometry} in the setting of smooth manifolds.  Thus a semi-conformal map generalizes this notion.  

Given a graph $\Ga = (V,E)$ and a vertex $x\in V$, then a function $f : V \ra \RR$ is harmonic at $x$ if $\Delta f(x) = 0$ -- we will call such a function a \emph{local harmonic function}.  In \cite{Ur-1, Ur-2} it is shown that a mapping between graphs pulls back local harmonic functions to local harmonic functions if and only if it is semi-conformal.  This concept is the discrete analogue of a \emph{harmonic morphism} \cite{Ba-Wo-3}. 

We now introduce one of the fundamental objects of our study, namely a \emph{holomorphic function} on a graph. 

\begin{definition}  Let $\Ga = (V,E)$ be a (not necessarily finite) graph, then a function $\phi : V\ra \CC$ is called {\rm holomorphic at} $x\in V$ if 
$$
\sum_{y\sim x} (\dd \phi (\ov{xy}))^2 = \sum_{y\sim x} (\phi (y) - \phi (x))^2 = 0\,.
$$
We say that $\phi : \Ga \ra \CC$ is {\rm holomorphic} if it is holomorphic at every vertex $x\in V$. 
\end{definition}

The notion is a natural adaptation of that of a semi-conformal mapping $\phi : M^m \ra \CC$ from a Riemannian $m$-manifold into the complex plane, as discussed in Section \ref{sec:SFR}.  For, $\phi : U \subset \RR^2 \ra \CC$ is semi-conformal if and only if
$$
\left( \frac{\pa \phi}{\pa x}\right)^2 + \left( \frac{\pa \phi}{\pa y}\right)^2 = 4 \frac{\pa \phi}{\pa z}\frac{\pa \phi}{\pa \ov{z}} = 0\,.
$$
That is, if and only if $\phi$ is holomorphic or anti-holomorphic.  But on a graph, we do not a priori have a notion of orientation, which in the plane is precisely what distinguishes holomorphic from anti-holomorphic, which justifies the above definition.  However, we do sacrifice \emph{linearity} in the equation for holomorphicity, which is an essential ingredient in the study by Baker and Norine who develop their theory using harmonic functions.

\medskip

\begin{center}
\setlength{\unitlength}{0.254mm}
\begin{picture}(244,92)(15,-156)
        \allinethickness{0.254mm}\path(40,-120)(120,-80) 
        \allinethickness{0.254mm}\path(120,-80)(240,-100) 
        \allinethickness{0.254mm}\path(240,-100)(160,-140) 
        \allinethickness{0.254mm}\path(160,-140)(40,-120) 
        \allinethickness{0.254mm}\path(120,-80)(200,-80) 
        \allinethickness{0.254mm}\path(200,-80)(240,-100) 
        \allinethickness{0.254mm}\path(160,-140)(80,-140) 
        \allinethickness{0.254mm}\path(80,-140)(40,-120) 
        \allinethickness{0.254mm}\path(240,-100)(210,-130) 
        \allinethickness{0.254mm}\path(210,-130)(160,-140) 
        \allinethickness{0.254mm}\path(40,-120)(70,-90) 
        \allinethickness{0.254mm}\path(70,-90)(120,-80) 
        \allinethickness{0.254mm}\special{sh 0.3}\put(200,-80){\ellipse{4}{4}} 
        \allinethickness{0.254mm}\special{sh 0.3}\put(240,-100){\ellipse{4}{4}} 
        \allinethickness{0.254mm}\special{sh 0.3}\put(210,-130){\ellipse{4}{4}} 
        \allinethickness{0.254mm}\special{sh 0.3}\put(160,-140){\ellipse{4}{4}} 
        \allinethickness{0.254mm}\special{sh 0.3}\put(80,-140){\ellipse{4}{4}} 
        \allinethickness{0.254mm}\special{sh 0.3}\put(40,-120){\ellipse{4}{4}} 
        \allinethickness{0.254mm}\special{sh 0.3}\put(70,-90){\ellipse{4}{4}} 
        \allinethickness{0.254mm}\special{sh 0.3}\put(120,-80){\ellipse{4}{4}} 
        \put(215,-141){\shortstack{$0$}} 
        \put(245,-101){\shortstack{$\ii$}} 
        \put(195,-76){\shortstack{$1+\ii$}} 
        \put(110,-76){\shortstack{$1+2\ii$}} 
        \put(53,-84){\shortstack{$2+2\ii$}} 
        \put(11,-114){\shortstack{$2+\ii$}} 
        \put(70,-156){\shortstack{$1+\ii$}} 
        \put(160,-156){\shortstack{$1$}} 
\end{picture}
\end{center}
\medskip
\begin{center}
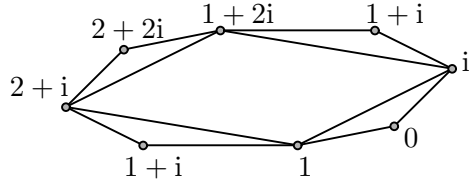
{\small Figure 1:  \emph{Example of a finite graph endowed with a holomorphic function}}
\end{center}
\medskip

We now prove an analogue in the context of holomorphic functions, of a theorem of Urakawa \cite{Ur-1, Ur-2}, that holomorphic (or semi-conformal) mappings between graphs are characterized as those mappings which preserve harmonic functions.

\begin{proposition} \label{prop:holo}
Let $\phi : \Ga_1 = (V_1, E_1) \ra (V_2, E_2)$ be a mapping between graphs.  Then $\phi$ is holomorphic if and only if it preserves local holomorphic functions, that is, if $f: V_2 \ra \CC$ is holomorphic at $\phi (x)$ $(x\in V_1$), then $f\circ \phi$ is holomorphic at $x$.  In particular, if $\phi : \Ga_1 \ra \Ga_2$ is holomorphic, then $f\circ \phi$ is also holomorphic for every holomorphic function $f : V_2 \ra \CC$.
\end{proposition}

\noindent \emph{Proof}:  Suppose that $\phi : \Ga_1 \ra \Ga_2$ is holomorphic and let $f: V_2 \ra \CC$ be holomorphic at $y \in V_2$.  Consider the function $f \circ \phi$.  We show that it is holomorphic at each point $x$ with $\phi (x) = y$.  Now
\begin{eqnarray*}
\sum_{x'\sim x}\Big( (f\circ \phi )(x') - (f \circ \phi )(x)\Big)^2 & = & \sum_{x'\sim x}(f(\phi (x')) - f(y))^2 \\
 & = & \la (x) \sum_{y'\sim y} (f(y')- f(y))^2 = 0 \,,
\end{eqnarray*}
by the holomorphicity of $\phi$.

Conversely, suppose that $\phi : \Ga_1 \ra \Ga_2$ preserves local holomorphic functions.  Let $y \in V_2$ and let $x \in \phi^{-1}(y) \in V_1$.  If there is only one vertex $y_1 \sim y$, then the condition of holomorphicity at $x$ is trivially satisfied, so we may suppose there are at least two distinct vertices joined by an edge to $y$.  Let $y_1, y_2 \sim y$.  We want to show that $\la (x, y_1) = \la (x, y_2)$.  Consider the function $f$ holomorphic at $y$ given by $f(y) = 0$, $f(y_1) = \ii$, $f(y_2) = 1$ and $f(y') = 0$ for all $y'\sim y$ with $y' \neq y_1, y_2$.  By hypothesis, $f\circ \phi$ is holomorphic at $x$, so that, if $x_1, \ldots , x_r \sim x$ satisfy $\phi (x_1) = \cdots = \phi (x_r) = y_1$ and $x_{r+1}, \ldots , x_{r+s} \sim x$ satisfy $\phi (x_{r+1}) = \cdots = \phi (x_{r+s}) = y_2$, then
$$
\sum_{x'\sim x} \Big((f\circ \phi )(x') - (f\circ \phi )(x)\Big)^2 = - r + s\,,
$$
which must vanish, so that $r = s$ and $\la (x, y_1 ) = \la (x, y_2)$.  Since $y_1, y_2 \sim y$ are arbitrarily chosen, we conclude that $\phi$ is holomorphic.  
\hfill q.e.d. 

\medskip

We will consider a pair $(\Ga , \phi )$, of a graph together with a holomorphic function $\phi : \Ga \ra \CC$, as a (static) field.  Later on, we will consider how to introduce a dynamic into the field.  It may be appropriate in the context of quantum field theory to view $\phi$ as a probability amplitude defined at each vertex.  Note that if $\phi : \Ga \ra \CC$ is a holomorphic function, then so is $c\phi + a$ for any complex constants $a,c \in \CC$.

A holomorphic function can be viewed as a special case of a more general object, which we refer to as an isotropic $1$-form.

\begin{definition} \label{def:isotropic-form}  Let $\om$ be a $1$-form defined on a graph $\Ga = (V,E)$.  Then we call $\om$ {\rm isotropic} if 
$$
\sum_{y\sim x}(\om (\ov{xy}))^2 = 0\,,
$$
at each vertex $x\in V$.
\end{definition}

Then the derivative $\dd\phi$ of a holomorphic function is an isotropic $1$-form.  Conversely, we require an integrability condition on $\om$ in order that it be the derivative of a function.  This amounts to the requirement that $\sum_k\om (e_k)$ should vanish around any cycle $\{ e_k\}_k$ (a cycle being a sequence of directed edges $\{ e_1, e_2, \ldots , e_r\}$ such that the point of arrival of $e_k$ is the start point of $e_{k+1}$ with $e_{r+1}$ then being identified with $e_1$).  For if this is the case, then we define $\phi$ at a fixed vertex $x_0$, say to take the value $\phi_0$ and then set $\phi (y) = \phi_0 + \om (\vec{xy})$ for $y\sim x$.  Continuation of this process to all vertices is well-defined on account of the cycle condition. 

 A quantum graph is a metric graph, such that each edge supports a solution to the $1$-dimensional Schr\"odinger equation with a compatibility condition at each vertex, see \cite{Gn-Sm} and the references cited therein.  We can view the pair $(\Ga , \phi )$ of a graph endowed with a holomorphic function as a similar structure, where we replace a solution to the $1$-dimensional Schr\"odinger equation on an edge $\vec{xy}$ by the amplitude $\phi (y) - \phi (x)$.  The compatibility condition at each vertex becomes $\sum_{y\sim x} (\phi (y) - \phi (x))^2 = 0$.  

A spin network, in its more recent formulation, consists of a graph where each edge has a label which corresponds to a representation of a particular group.  To each vertex is associated an intertwiner which relates these different representations.  The original spin networks of Penrose consist of regular graphs with each vertex having degree $3$ and with associated group $\SU (2)$ \cite{Ro-Sm}.  Note that the character of an irreducible representation is an algebraic integer, that is, it is the root of some monic equation.  We do not know if there may be a deeper connection between spin networks and pairs $(\Ga , \om )$, where $\om$ is an isotropic $1$-form with the different values $\om (\vec{xy})$ corresponding to characters of representations satisfying polynomial identities at each vertex.   

Another interesting construction is the following.  Given $n$ complex numbers $z_1, z_2, \ldots , z_n$ satisfying $\sum_{k = 1}^n z_k{}^2 = 0$, then one can construct an $n$-dimensional cube in $\RR^n$ such that there exists an orthogonal projection from $\RR^n$ onto $\CC$ which maps one vertex $v_0\in \RR^n$ to $0\in \CC$ and its neighbouring vertices $v_1, v_2, \ldots , v_n \in \RR^n$ to the points $z_1, z_2, \ldots , z_n$.  Conversely, given any orthogonal projection $\pi :\RR^n \ra \CC$, then the complex numbers $z_k = \pi (v_k-v_0)$ satisfy $\sum z_k{}^2 = 0$.  This property is known under the name of \emph{Gauss' fundamental theorem of axonometry} \cite{Ga} and when $n = 3$, the projection of the vertices is known as \emph{orthographic projection}.  For example, the three dimensional cube supports the holomorphic function indicated in Figure 2.

\medskip

\begin{center}

\setlength{\unitlength}{0.254mm}
\begin{picture}(180,167)(25,-176)
        \allinethickness{0.254mm}\path(40,-40)(80,-80) 
        \allinethickness{0.254mm}\path(80,-80)(160,-60) 
        \allinethickness{0.254mm}\path(160,-60)(120,-20) 
        \allinethickness{0.254mm}\path(120,-20)(40,-40) 
        \allinethickness{0.254mm}\path(80,-80)(80,-160) 
        \allinethickness{0.254mm}\path(160,-60)(160,-140) 
        \allinethickness{0.254mm}\path(160,-140)(80,-160) 
        \allinethickness{0.254mm}\path(40,-40)(40,-120) 
        \allinethickness{0.254mm}\path(40,-120)(80,-160) 
        \allinethickness{0.254mm}\path(160,-140)(120,-100) 
        \allinethickness{0.254mm}\path(120,-100)(120,-75) 
        \allinethickness{0.254mm}\path(120,-60)(120,-20) 
        \allinethickness{0.254mm}\path(40,-120)(75,-110) 
        \allinethickness{0.254mm}\path(120,-100)(95,-105) 
        \allinethickness{0.254mm}\special{sh 0.3}\put(160,-60){\ellipse{4}{4}} 
        \allinethickness{0.254mm}\special{sh 0.3}\put(160,-140){\ellipse{4}{4}} 
        \allinethickness{0.254mm}\special{sh 0.3}\put(120,-100){\ellipse{4}{4}} 
        \allinethickness{0.254mm}\special{sh 0.3}\put(120,-20){\ellipse{4}{4}} 
        \allinethickness{0.254mm}\special{sh 0.3}\put(40,-40){\ellipse{4}{4}} 
        \allinethickness{0.254mm}\special{sh 0.3}\put(80,-80){\ellipse{4}{4}} 
        \allinethickness{0.254mm}\special{sh 0.3}\put(40,-120){\ellipse{4}{4}} 
        \allinethickness{0.254mm}\special{sh 0.3}\put(80,-160){\ellipse{4}{4}} 
        \put(165,-156){\shortstack{$-1$}} 
        \put(80,-176){\shortstack{$0$}} 
        \put(25,-126){\shortstack{$1$}} 
        \put(130,-101){\shortstack{$0$}} 
        \put(170,-66){\shortstack{$-1+\sqrt{2}\ii$}} 
        \put(130,-21){\shortstack{$\sqrt{2}\ii$}} 
        \put(80,-71){\shortstack{$\sqrt{2}\ii$}} 
        \put(24,-29){\shortstack{$1+\sqrt{2}\ii$}} 
\end{picture}
\end{center}
\medskip
\begin{center}
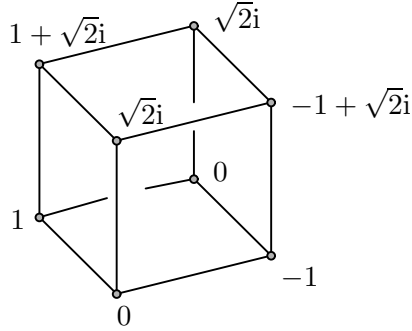
{\small Figure 2:  \emph{The $1$-skeleton of the cube endowed with a holomorphic function}}
\end{center}
\medskip 

The projection of the vertices of other regular polyhedra, satisfy other polynomial equations.  For example, the equation $(z_1 + \cdots z_n)^2 - (n+1)(z_1{}^2 + \cdots + z_n{}^2) = 0$ is satisfied by the orthogonal projections $z_1, \ldots , z_n$ of the vertices of a regular tetrahedron \cite{Ea-Pe}. 

The case of the cube  shows how we can see an $n$-dimensional space arising from a regular graph $(\Ga , \phi )$ with common vertex degree $n$ endowed with a holomorphic function.  Specifically, at each vertex $x$, the complex numbers $\phi (y) - \phi (x)$ $(y\sim x)$ generate a cube in $\RR^n$.   

On some infinite graphs, the construction of a holomorphic function can be easily achieved.  For example, let $\Ga$ be the integer lattice in $\RR^N$, with edges joining vertices whose components differ by $1$ in a single entry.  Then given any complex valued function $g_0$ defined on the set $\{ (x_1, x_2, \ldots , x_{N-1}, 0)\in \ZZ^N\}$ and another one $g_1$ defined on $\{ (x_1, x_2, \ldots , x_{N-1}, 1)\in \ZZ^N\}$, we can now construct a holomorphic function $\phi$ by extension.  Explicitly, $\phi (x_1, x_2, \ldots , x_{N-1}, 2)$ is obtained by solving the equation
\begin{eqnarray*}
\sum_{k = 1}^{N-1} \left\{ \big( g_1 (x_1, \ldots x_k - 1, \ldots , x_{N-1}, 1) - g_1 (x_1, \ldots x_k, \ldots , x_{N-1}, 1)\big)^2\right\} \\
+ \big( g_0 (x_1,  \ldots , x_{N-1}, 0) - (g_1 (x_1, \ldots , x_{N-1}, 1)\big)^2 \qquad \\ + \big(\phi (x_1,  \ldots , x_{N-1}, 2) - (g_1 (x_1, \ldots , x_{N-1}, 1)\big)^2 = 0
\end{eqnarray*}
for $\phi (x_1,  \ldots , x_{N-1}, 2)$, and so on.  In general, at each step there will be two solutions and so infinitely many branches will be defined on $\RR^N$.  We can view such holomorphic functions as solving an intitial value problem: given a function $g$ and its normal derivative on a hypersurface $S$, find a holomorphic function $\phi$ which coincides with $g$ and has the same normal derivative on $S$.  However, finding \emph{finite} graphs which support a holomorphic function seems much harder and at present, using a computer, we can only test examples with a small number of vertices.  For example, MAPLE fails to find a holomorphic function on the $1$-skeleton of the dodecahedron in a reasonable time, however, it does show the existence of isotropic $1$-forms.   

We now wish to show how, given a graph endowed with a holomorphic function, we can recover a spinor field on the graph.  Let $\Ga = (V,E)$ be a regular graph with common vertex degree $m$.  An \emph{orientation on} $\Ga$ is a colouring of the edges of the graph with the numbers $1, 2, \ldots , m$.  By a \emph{colouring}, we mean an assignment of a number $k \in \{ 1, 2, \ldots , m\}$ to each edge so that no two edges incident with the same vertex have the same colour.  For example, the $1$-skeleton of the cube, above, is coloured as follows:  

\medskip

\begin{center}

\setlength{\unitlength}{0.254mm}
\begin{picture}(168,152)(20,-166)
        \allinethickness{0.254mm}\path(40,-40)(80,-80) 
        \allinethickness{0.254mm}\path(80,-80)(160,-60) 
        \allinethickness{0.254mm}\path(160,-60)(120,-20) 
        \allinethickness{0.254mm}\path(120,-20)(40,-40) 
        \allinethickness{0.254mm}\path(80,-80)(80,-160) 
        \allinethickness{0.254mm}\path(40,-40)(40,-120) 
        \allinethickness{0.254mm}\path(40,-120)(80,-160) 
        \allinethickness{0.254mm}\path(80,-160)(160,-140) 
        \allinethickness{0.254mm}\path(160,-140)(160,-60) 
        \allinethickness{0.254mm}\path(160,-140)(120,-100) 
        \allinethickness{0.254mm}\path(120,-100)(120,-75) 
        \allinethickness{0.254mm}\path(120,-60)(120,-20) 
        \allinethickness{0.254mm}\path(120,-100)(95,-105) 
        \allinethickness{0.254mm}\path(75,-110)(40,-120) 
        \put(125,-166){\shortstack{$1$}} 
        \put(170,-106){\shortstack{$2$}} 
        \put(85,-126){\shortstack{$2$}} 
        \put(65,-26){\shortstack{$3$}} 
        \put(100,-71){\shortstack{$3$}} 
        \put(20,-86){\shortstack{$2$}} 
        \put(45,-151){\shortstack{$3$}} 
        \put(65,-61){\shortstack{$1$}} 
        \put(145,-41){\shortstack{$1$}} 
\end{picture}
\medskip
\begin{center}
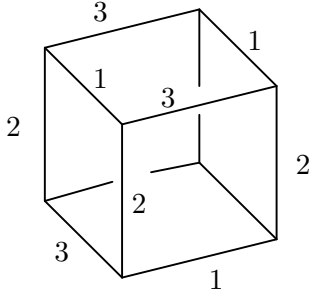
{\small Figure 3:  \emph{The $1$-skeleton of the cube with an orientation giving colouring}}
\end{center}
\medskip
\end{center}
   
Let $\Ga = (V,E)$ be a regular graph of degree $3$ which is oriented by the colours $\{ 1,2,3\}$.  Suppose further, that $\Ga$ is endowed with an isotropic $1$-form $\om$.  Then, given a vertex $x\in V$, we can associate to $x$ a triple of complex numbers $\xi (x) = (\xi_1, \xi_2, \xi_3)$, where $x\sim y_1,y_2,y_3$, $\xi_k = \om (\vec{xy_k})$ and we suppose the edge $\ov{xy_k}$ has colour $k$ $(k = 1,2,3)$.  Since $\xi_1{}^2 + \xi_2{}^2 + \xi_3{}^2 = 0$, the symmetric matrix
$$
(\Om^{AB}):= \left( \begin{array}{cc} - \xi_2 - \xi_3\ii & \xi_3 \\ \xi_3 & \xi_2 - \xi_3\ii 
\end{array}
\right) \quad (A,B \in \{ 0,1\} )
$$
has determinant zero and so can be written in the form $\Om^{AB} = \mu^A\mu^B$, for some spinor $(\mu^A) \in \CC^2$ (defined up to sign).  We therefore have a spinor field $\mu^A$ on $\Ga$ that provides the analogue of the spinor field on $\RR^3$ which generates an SFR in Minkowski space, as described in Section \ref{sec:SFR}. 

We can proceed further and construct the analogue of the vector field $U$ (tangent to the associated conformal foliation in the smooth case) at each vertex.  In fact, $\mu = \mu^0/\mu^1 = - (\xi_2 + \ii \xi_3)/\xi_1$ represents the direction of $U$ in the chart given by stereographic projection, so that
$$
U = \frac{1}{|\xi_1|^2 + |\xi_2 + \ii \xi_3|^2}\left( |\xi_2 + \ii \xi_3|^2 - | \xi_1|^2, - \ov{\xi_1}(\xi_2 + \ii \xi_3)\right)\,.
$$
It is now possible to consider the discrete analogue of equation (\ref{evol-scm}):
$$
\dd \left( \frac{\pa\phi_n}{\pa n}\right) (U) = - \Delta \phi_n\,,
$$
for a family of complex-valued functions $\{ \phi_n\}$ parametrized by the natural numbers, equivalently:
\begin{equation} \label{evol-scm-discrete}
\dd \phi_{n+1}(U) = - \Delta \phi_n\,.
\end{equation}
However, care needs to be taken in the choice of sign of $U$ when applying this equation, since our construction has essentially only found a \emph{non-oriented} direction $U$ at each vertex.  In the case when $\phi_n$ is a given \emph{holomorphic} function, we can ask whether (\ref{evol-scm-discrete}) determines successive functions $\phi_{n+1}$ which are also holomorphic.  We do not have a general result to this effect, but it does turn out to be the case for the graph consisting of the $1$-skeleton of the cube.   The following table constructs the successive holomorphic function, which is unique up to addition of a constant.

\medskip

\begin{center}
{\scriptsize
\begin{tabular}{|c|c|c|c|c|c|}  \hline
Vertex & $\xi$ & $U$ & $\dd \phi_{n+1}(U)$ & $\Delta \phi_n$ & $\phi_{n+1}$ \\ \hline
$1$ &  $(1, \sqrt{2}\ii , 1)$ & $\frac{1}{\sqrt{2}}(1,0,-1)$ & $\frac{1}{\sqrt{2}}(\phi_{n+1}(2) - \phi_{n+1}(7))$ & $- \frac{\sqrt{2}}{3}(\sqrt{2} + \ii )$ & $\frac{2\sqrt{2}}{3}$  \\ 
$2$ &  $(-1, \sqrt{2}\ii , -1)$ & $\frac{1}{\sqrt{2}}(1,0,-1)$ & $\frac{1}{\sqrt{2}}(\phi_{n+1}(1) - \phi_{n+1}(8))$ & $ \frac{\sqrt{2}}{3}(\sqrt{2} - \ii )$ & $\frac{2(\sqrt{2}+\ii )}{3}$  \\ 
$3$ &  $(1, -\sqrt{2}\ii , 1)$ & $\frac{1}{\sqrt{2}}(1,0,-1)$ & $\frac{1}{\sqrt{2}}(\phi_{n+1}(5) - \phi_{n+1}(4))$ & $- \frac{\sqrt{2}}{3}(\sqrt{2} - \ii )$ & $\frac{2(\sqrt{2}+ \ii )}{3}$  \\ 
$4$ &  $-(1, \sqrt{2}\ii , 1)$ & $-\frac{1}{\sqrt{2}}(1,0,-1)$ & $\frac{1}{\sqrt{2}}(\phi_{n+1}(3) - \phi_{n+1}(6))$ & $ \frac{\sqrt{2}}{3}(\sqrt{2} + \ii )$ & $\frac{2\sqrt{2}}{3}$  \\ 
$5$ &  $-(1, \sqrt{2}\ii , 1)$ & $-\frac{1}{\sqrt{2}}(1,0,-1)$ & $\frac{1}{\sqrt{2}}(\phi_{n+1}(3) - \phi_{n+1}(6))$ & $ \frac{\sqrt{2}}{3}(\sqrt{2} + \ii )$ & $\frac{2\ii}{3}$  \\ 
$6$ &  $(1, -\sqrt{2}\ii , 1)$ & $-\frac{1}{\sqrt{2}}(1,0,-1)$ & $\frac{1}{\sqrt{2}}(\phi_{n+1}(5) - \phi_{n+1}(4))$ & $ -\frac{\sqrt{2}}{3}(\sqrt{2} - \ii )$ & $0$  \\ 
$7$ &  $(-1, \sqrt{2}\ii , -1)$ & $-\frac{1}{\sqrt{2}}(1,0,-1)$ & $\frac{1}{\sqrt{2}}(\phi_{n+1}(1) - \phi_{n+1}(8))$ & $ \frac{\sqrt{2}}{3}(\sqrt{2} - \ii )$ & $0$  \\ 
$8$ &  $(1, \sqrt{2}\ii , 1)$ & $-\frac{1}{\sqrt{2}}(1,0,-1)$ & $\frac{1}{\sqrt{2}}(\phi_{n+1}(2) - \phi_{n+1}(7))$ & $ -\frac{\sqrt{2}}{3}(\sqrt{2} + \ii )$ & $\frac{2\ii}{3}$  \\  \\ \hline
\end{tabular}
}
\end{center}

\medskip

\section{The twistor correspondence between graphs} \label{sec:line-graph} 
Twistor space, as first introduced by R. Penrose \cite{Pe-1}, is the space whose points correspond to light rays in Minkowski space.  More precisely, there is a $5$-real dimensional CR-submanifold of $\CP^3$ whose points are the light rays.  In order to complete the picture it is necessary to compactify and to complexify $\MM^4$ to the complex Grassmannian $G_2(\CC^4)$ of complex $2$-planes through the origin in $\CC^4$.  Via the twistor double fibration, a point of $\CP^3$ now determines an $\al$-plane in $G_2(\CC^4)$, which, if it intersects the real space $\MM^4$, does so in a null geodesic (see, for example \cite{Wa-We}. 

On the other hand, associated to the three-dimensional space forms is their \emph{mini-twistor space}: the space of all geodesics.  For example, the mini-twistor space of $\RR^3$ is the complex surface given by the tangent bundle to the $2$-sphere: $TS^2$; each line in $\RR^3$ being defined by its direction $u \in S^2$ and its displacement from the origin $c \in T_uS^2$ ($c$ is the unique vector starting at the origin which hits the line at right angles) (see, \cite{Ba-Wo-3}). In view of these correspondences, it is very natural to define the twistor dual of a graph to be the graph whose vertices are the edges of the original graph, where two vertices are connected if and only if the corresponding edges in the original graph are incident.  This dual graph is a well-known classical concept called the line-graph.

Precisely, given a graph $\Ga = (V,E)$, then the \emph{line-graph} or \emph{twistor dual of} $\Ga$ is the graph $L_{\Ga } = (E,T)$, where, for $X,Y \in E$, we have $X\sim Y$ if and only if $X$ and $Y$ are incident in $\Ga$. The only connected graph that is isomorphic to its line-graph is a cyclic graph and H. Whitney showed that, with the exception of the graphs $K_3$ (the complete graph on three vertices) and $K_{1,3}$ (the bipartite graph with edges joining one vertex to three other unconnected vertices)), any two connected graphs with isomorphic line graphs are isomorphic \cite{Wh}.  Not every graph arises as the line-graph of a graph, specifically, there are nine classified graphs, such that provided a given graph $L$ doesn't contain one of them as a subgraph, then $L = L_{\Ga}$ is the line-graph of some graph $\Ga$ \cite{Va-Wi, Be}. As an example, Figure 4 shows the line-graph of the graph of Figure 1. 

\medskip

\begin{center}

\setlength{\unitlength}{0.254mm}
\begin{picture}(194,124)(43,-162)
        \allinethickness{0.254mm}\path(80,-60)(60,-140) 
        \allinethickness{0.254mm}\path(60,-140)(200,-140) 
        \allinethickness{0.254mm}\path(200,-140)(220,-60) 
        \allinethickness{0.254mm}\path(220,-60)(80,-60) 
        \allinethickness{0.254mm}\path(130,-40)(175,-40) 
        \allinethickness{0.254mm}\path(175,-40)(220,-60) 
        \allinethickness{0.254mm}\path(130,-40)(80,-60) 
        \allinethickness{0.254mm}\path(80,-60)(55,-85) 
        \allinethickness{0.254mm}\path(55,-85)(45,-120) 
        \allinethickness{0.254mm}\path(45,-120)(60,-140) 
        \allinethickness{0.254mm}\path(60,-140)(100,-160) 
        \allinethickness{0.254mm}\path(100,-160)(150,-160) 
        \allinethickness{0.254mm}\path(150,-160)(200,-140) 
        \allinethickness{0.254mm}\path(200,-140)(225,-115) 
        \allinethickness{0.254mm}\path(225,-115)(235,-80) 
        \allinethickness{0.254mm}\path(235,-80)(220,-60) 
        \allinethickness{0.254mm}\special{sh 0.3}\put(200,-140){\ellipse{4}{4}} 
        \allinethickness{0.254mm}\special{sh 0.3}\put(225,-115){\ellipse{4}{4}} 
        \allinethickness{0.254mm}\special{sh 0.3}\put(235,-80){\ellipse{4}{4}} 
        \allinethickness{0.254mm}\special{sh 0.3}\put(220,-60){\ellipse{4}{4}} 
        \allinethickness{0.254mm}\special{sh 0.3}\put(175,-40){\ellipse{4}{4}} 
        \allinethickness{0.254mm}\special{sh 0.3}\put(130,-40){\ellipse{4}{4}} 
        \allinethickness{0.254mm}\special{sh 0.3}\put(80,-60){\ellipse{4}{4}} 
        \allinethickness{0.254mm}\special{sh 0.3}\put(55,-85){\ellipse{4}{4}} 
        \allinethickness{0.254mm}\special{sh 0.3}\put(45,-120){\ellipse{4}{4}} 
        \allinethickness{0.254mm}\special{sh 0.3}\put(60,-140){\ellipse{4}{4}} 
        \allinethickness{0.254mm}\special{sh 0.3}\put(100,-160){\ellipse{4}{4}} 
        \allinethickness{0.254mm}\special{sh 0.3}\put(150,-160){\ellipse{4}{4}} 
        \allinethickness{0.254mm}\path(150,-160)(175,-145) 
        \allinethickness{0.254mm}\path(225,-115)(210,-125) 
        \allinethickness{0.254mm}\path(200,-130)(190,-135) 
        \allinethickness{0.254mm}\path(235,-80)(215,-115) 
        \allinethickness{0.254mm}\path(220,-60)(225,-90) 
        \allinethickness{0.254mm}\path(175,-40)(200,-55) 
        \allinethickness{0.254mm}\path(225,-75)(235,-80) 
        \allinethickness{0.254mm}\path(130,-40)(185,-50) 
        \allinethickness{0.254mm}\path(80,-60)(145,-45) 
        \allinethickness{0.254mm}\path(55,-85)(75,-70) 
        \allinethickness{0.254mm}\path(110,-50)(130,-40) 
        \allinethickness{0.254mm}\path(45,-120)(65,-85) 
        \allinethickness{0.254mm}\path(60,-140)(55,-110) 
        \allinethickness{0.254mm}\path(100,-160)(75,-145) 
        \allinethickness{0.254mm}\path(45,-120)(55,-130) 
        \allinethickness{0.254mm}\path(60,-140)(150,-160) 
        \allinethickness{0.254mm}\path(100,-160)(120,-155) 
        \allinethickness{0.254mm}\path(135,-150)(165,-145) 
\end{picture}

\end{center}

\medskip
\begin{center}
{\small Figure 4:  \emph{The line-graph of the graph of Figure 1}}
\end{center}
\medskip

We can now pursue the twistor correspondence, so that a vertex of a graph $\Ga$ corresponds to a complete subgraph of the line graph $L_{\Ga}$.  This latter object is then the discrete analogue of the complex projective line corresponding to all the light rays passing through a given point.  If now $\Ga$ is endowed with an isotropic $1$-form $\om : T\Ga \ra \CC$, then, on giving each edge a direction, we can define a corresponding \emph{dual} function $\psi : V(L_{\Ga}) \ra \CC$, by $\psi (X) = \om (\vec{xy})$, where $X = \ov{xy}$ has direction $\vec{xy}$.  It follows that if $x \in V(\Ga )$ and $C_x$ is the complete subgraph of $L_{\Ga}$ corresponding to $x$, then
\begin{equation} \label{psi}
\sum_{X\in C_x} \psi (X)^2 = 0\,.
\end{equation}   
Note that this latter condition is independent of the choice of direction given to each edge in $\Ga$.

Conversely, given a graph $L$ which is the line graph of a graph $\Ga$, and a function $\psi : V(L) \ra \CC$ satisfying $\eqref{psi}$ for each complete subgraph $C_x$ corresponding to a vertex $x \in V(\Ga )$, then on giving each edge in $\Ga$ a direction, we can define an isotropic $1$-form on $\Ga$.  If further $\Ga$ is regular of degree three and oriented by colour, as described in the previous section, we then have a spinor field $\mu^A$ on $\Ga$ giving the analogue of an SFR.  This provides a discrete analogue of the Kerr Theorem, which associates to a complex analytic surface in $\CP^3$, a shear-free ray congruence in Minkowski space.

\end{document}